\documentclass[journal]{IEEEtran}
\usepackage{amsmath,amsfonts}
\usepackage{algorithmic}
\usepackage{array}
\usepackage{textcomp}
\usepackage{stfloats}
\usepackage{url}
\usepackage{verbatim}
\usepackage{graphicx}
\usepackage{balance}
\usepackage[numbers,sort&compress]{natbib}
\usepackage[colorlinks,
            linkcolor=red,
            anchorcolor=blue,
            citecolor=green]{hyperref}
            
\usepackage{array}
\usepackage{multirow}
\usepackage{longtable}
\usepackage{rotating}
\usepackage{booktabs}
\usepackage{float}  
\usepackage{subfigure}

\usepackage{footnote}
 \usepackage{threeparttable}
 \usepackage{flushend}
 \usepackage{amssymb}
\def\BibTeX{{\rm B\kern-.05em{\sc i\kern-.025em b}\kern-.08em
             T\kern-.1667em\lower.7ex\hbox{E}\kern-.125emX}}
\begin{document}
%%%%%%%%%%%%%%%%%%%%%%%%%%%%%%%
\title{Using Geometrical information to Measure the Vibration of  A Swaying Millimeter-wave Radar}
  \author{Chengyao Tang, Yongpeng Dai, Zhi Li, Tian Jin, \IEEEmembership{Member, IEEE}

				  \thanks{Manuscript created Nov. 1, 2023. This work was supported by the National Natural Science Foundation of China (Grant Nos. 61971430). (\textit{Corresponding author: Tian Jin}).}
				  \thanks{Chengyao Tang, Yongpeng Dai, Zhi Li and Tian Jin are with the College of Electronic Science and Technology, National University of Defense Technology, Changsha 410073, China. \textit{(E-mail: \{cyt, dai\_yongpeng, lizhi, songyongping08, tianjin\}@nudt.edu.cn)}.
				  }% 
         }

  \markboth{Journal,~Vol.~18, No.~9, December~2023}%
  {How to Use the IEEEtran \LaTeX \ Templates} 

  \maketitle

\begin{abstract}
This paper presents two new, simple yet effective approaches to measure the vibration of a swaying millimeter-wave radar (mmRadar) utilizing geometrical information. Specifically, for the planar vibrations, we firstly establish an equation based on the area difference between the swaying mmRadar and the reference objects at different moments, which enables the quantification of planar displacement. Secondly, volume differences are also utilized with the same idea, achieving the self-vibration measurement of a swaying mmRadar for spatial vibrations. Experimental results confirm the effectiveness of our methods, demonstrating its capability to estimate both the amplitude and a crude direction of the mmRadar's self-vibration.
\end{abstract}

%%%%%%%%%%%%%%%%%%%%%%%%%%%%%%%%%%%%%%%%%%%%%%%%%%%%%%%%%%%%%%%%%%%%%%%%%%%%%%%%%%%%%%%%%%%%%%%
\graphicspath{Figures}
%%%%%%%%%%%%%%%%%%%%%%%%%%%%%%%%%%%%%%%%%%%%%%%%%%%%%%%%%%%%%%%%%%%%%%%%%%%%%%%%%%%%%%%%%%%%%%%
\section{Introduction}\label{Introduction}
\begin{figure}[t]
	\centering
	\includegraphics[width=0.88\columnwidth]{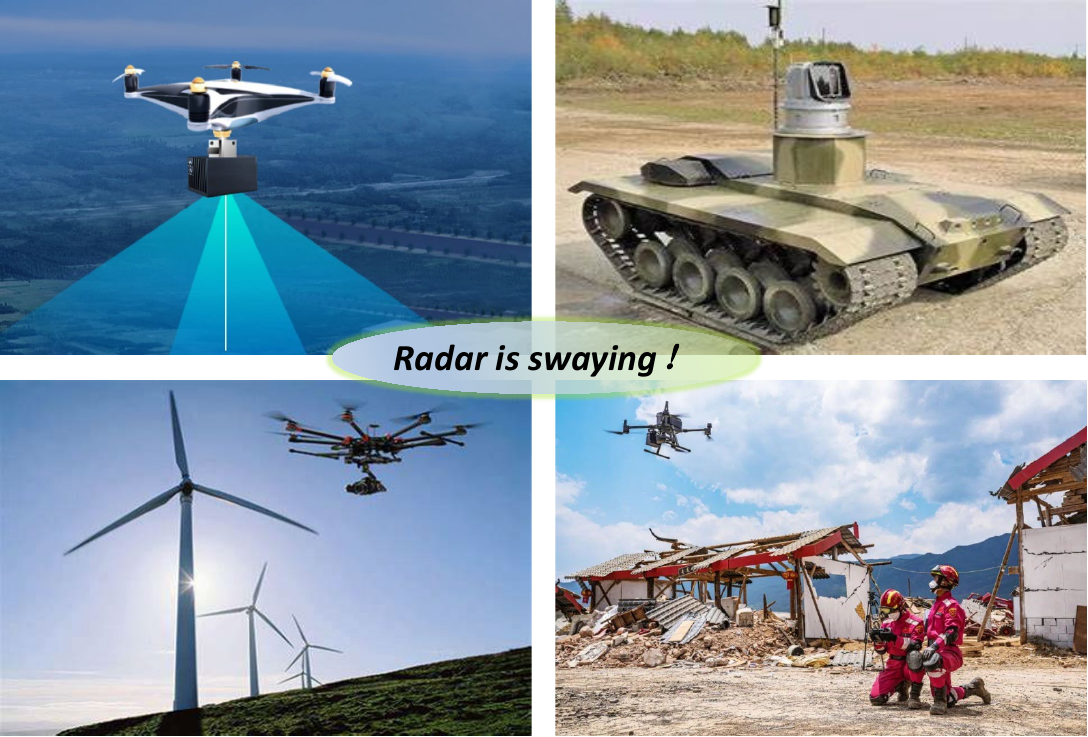}
	\caption{Typical application scenarios where swaying mmRadars exist such as battlefield environment monitoring, wind turbine inspection, and post-disaster aerial search and rescue. \label{examples}}
\end{figure}

\IEEEPARstart{T}{he} use of mobile platforms (such as drones or unmanned vehicles) equipped with millimeter-wave radar (mmRadar) for precise sensing tasks is gaining increasing attention. However, when these mobile platforms hover in the air or pause on the ground, they are still subject to influences such as wind or engine vibrations, causing the borne mmRadar to sway \cite{rongNoncontactVitalSign2021}, as shown in Fig. \ref{examples}. Clearly, measuring the vibration of a swaying mmRadar is an intriguing and meaningful task. On the one hand, we can use the vibration data obtained to derive current wind conditions or engine vibration information, which may be beneficial for meteorological observation and engine fault diagnostics. On the other hand, if these vibrations are considered undesirable and need to be cancelled, measuring the vibration information in advance can lay the groundwork for developing cancellation algorithms.
%%%%%%======================================%%%%%%%%%%%%%%%%
%%%%%%======================================%%%%%%%%%%%%%%%%
\par
However, most research involving the swaying radar scenarios has traditionally viewed the swaying state as detrimental, focusing more on how to directly suppress this self-vibration effect. Consequently, there is a scarcity of dedicated studies on the task of self-vibration measurement of a swaying radar. Nowadays, there are roughly three main measurement approaches. The first involves sensor-assisted measurement, such as using inertial guidance devices \cite{nakata2016RFTechniquesMotion,stockel2023CorrelationbasedMotionEstimation,Grisot2023Monitoring}. The second deploys an additional radar mounted on the other side of the mobile platform to measure its vibrations through dual radar echoes \cite{islam2021AdaptiveFilterTechnique,nakata2018MotionCompensationUnmanned}. The third approach achieves radar self-vibration measurement from a signal correlation perspective \cite{cardillo2021VitalSignDetection,stockel2023CorrelationbasedMotionEstimation,rong2021CardiacRespiratorySensing,rohman2021ThroughtheWallHumanRespiration,zhu2018FundamentalandHarmonicDualFrequencyDoppler,rahman2015signal}. While these methods are effective, both sensor-assisted and multi-radar systems add extra payload to the platforms, and synchronizing multiple devices poses a significant challenge. Moreover, the vibration of a swaying radar generally exhibit small amplitudes, and is a weak signal. Extracting vibration amplitudes using traditional signal processing techniques requires prolonged accumulation and computation time. Therefore, developing a measurement approach that is straightforward and does not require additional sensors would be highly advantageous.
%%%%%%======================================%%%%%%%%%%%%%%%%
\par
To this end, we propose two approaches to measure the vibration of a swaying mmRadar by the mmRadar itself, separately for planar and spatial vibrations. Our approaches eliminate the need for additional sensors and avoid complex synchronization operations. We employ a novel perspective—geometrical information—to achieve the self-vibration measurement with mmRadar. The origin of this interesting idea is outlined as follows.
%%%%%%%%%%%%%%%%%%%%%%%%%%%%%%%%%%%%%%%%%%%%%%%%%%%%%%%%%
\section{Inspiration}
Leveraging high-dimensional information to perform low-dimensional tasks is a proven effective solution. Considering that the measurement of radar self-vibration essentially involves observing displacement, an idea occurred to us: could we estimate this displacement through observations in higher dimensions? For instance, using information such as area and volume.
%%%%%%%%%%%%%%%%%%%%%%%%%%%%%%%%%%%%%%%%%%%%%%%%%%%%%%%%%
\section{Area-based measurement}
\subsection{Swaying Types}
\par
As a two-dimensional measure, area reflects the size of a finite planar region and is applicable solely to addressing planar-related issues.
\par
\begin{figure*}[t]
\centering
%%%%%%%%%%%%%%%%%%%%%%%%%%
    \subfigure[\centering{X-swaying sate}]{\label{swaying1}\includegraphics[height=5.5cm]{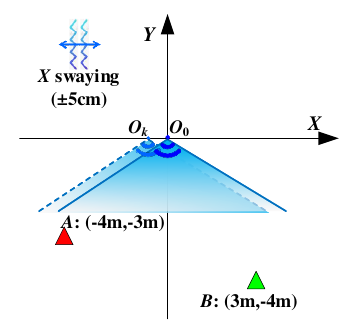}}
    \subfigure[\centering{Y-swaying sate}]{\label{swaying2}\includegraphics[height=5.5cm]{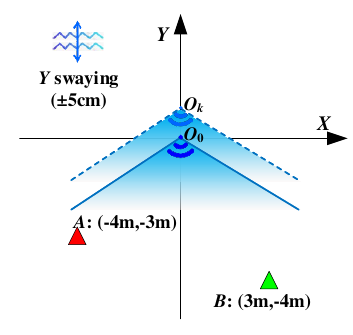}}
    \subfigure[\centering{XY-swaying sate}]{\label{swaying3}\includegraphics[height=5.5cm]{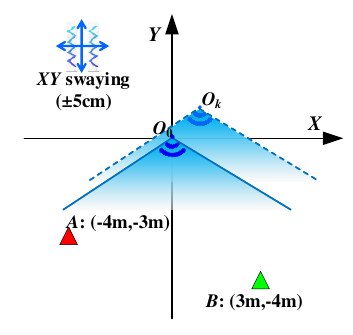}}
%%%%%%%%%%%%%%%%%%%%%%%%%%
\caption{Typical types of planar vibration for the swaying mmRadar.\label{FigXY}}
\end{figure*}

Therefore, we may start with the simple case of planar vibrations, where a radar swaying with the platform only oscillates within the plane. Within this plane, we establish an XY Cartesian coordinate system. To effectively utilize area information, let's assume there are two prominent targets within the plane (typically easy to identify), which serve as reference points. Based on the primary swaying direction, the radar's vibration can be categorized into three types: X-swaying, Y-swaying, and XY-swaying, as shown in Fig. \ref{FigXY}.
%%%%%%======================================%%%%%%%%%%%%%%%%
\par
Due to the third type, XY-swaying, is a composite state of the previous two, we will therefore proceed to establish separate models of X-swaying and Y-Swaying for analysis.
%%%%%%%%%%%%%%%%%%%%%%%%%%%%%%%%%%%%%%%%%%%%%%%%%%
\begin{figure*}[t]
\centering
%%%%%%%%%%%%%%%%%%%%%%%%%%
    \subfigure[\centering{X-swaying model}]{\label{xmodel}\includegraphics[height=7.5cm]{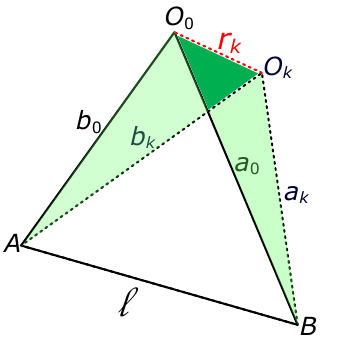}}
    \subfigure[\centering{Y-swaying model}]{\label{ymodel}\includegraphics[height=7.5cm]{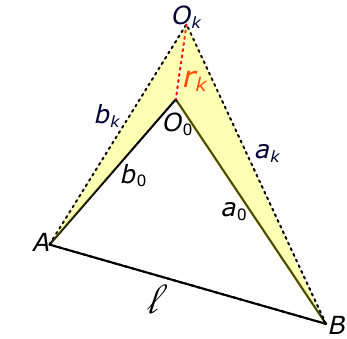}}
%%%%%%%%%%%%%%%%%%%%%%%%%%
\caption{The models leveraging area information for planar-vibration measurement.
\label{Fig1}}
\end{figure*}
%%%%%%======================================%%%%%%%%%%%%%%%%
%%%%%%======================================%%%%%%%%%%%%%%%%
%%%%%%======================================%%%%%%%%%%%%%%%%
%%%%%%======================================%%%%%%%%%%%%%%%%
\subsection{X-swaying model} 
In the X-swaying state, the radar mainly exhibits left-right vibration. The area information can be utilized to establish a geometric model as shown in Fig. \ref{xmodel}. Accordingly, the equation about the radar self-vibration value $r_k$ can be obtained as shown in Eq. (\ref{eqS1}).
%%%%%%%%%%%%%%%%%%%%%%%%%%%%%%%%%%%%%%%%%%%%%%%%%%%%%%%%%%%%%%%%%%
\begin{equation}
    \label{eqS1}
    \varDelta S=S_a\left( r_k \right) -S_b\left( r_k \right) .
\end{equation}
%%%%%%%%%%%%%%%%%%%%%%%%%%%%%%%%%%%%%%%%%%%%%%%%%%%%%%%%%%%%%%%%%%
\par
And the rough direction of the vibration can be determined by (\ref{eqD1}).
%%%%%%%%%%%%%%%%%%%%%%%%%%%%%%%%%%%%%%%%%%%%%%%%%%%%%%%%%%%%%%%%%%
\begin{equation}
    \label{eqD1}
    Direction=\left\{ \begin{matrix}
	Right:&		\varDelta \cos A\,\,>0 \&\& \varDelta \cos B\,\,<0\\
	Left:&		\varDelta \cos A\,\,<0 \&\& \varDelta \cos B\,\,>0\\
\end{matrix} \right..
\end{equation}
%%%%%%%%%%%%%%%%%%%%%%%%%%%%%%%%%%%%%%%%%%%%%%%%%%%%%%%%%%%%%%%%%%
\subsection{Y-swaying model} 
%%%%%%%%%%%%%%%%%%%%%%%%%%%%%%%%%%%%%%%%%%%%%%%%%%%%%%%%%%%%%%%%%%
Similarly, the model for the Y-swaying state can be obtained as shown in Fig. \ref{ymodel}. The measurement about the self-vibration r value $r_k$ can be estimated via Eq. (\ref{eqS1}).
\begin{equation}
    \label{eqS2}
    \varDelta S=S_a\left( r_k \right) +S_b\left( r_k \right)  .
\end{equation}
%%%%%%%%%%%%%%%%%%%%%%%%%%%%%%%%%%%%%%%%%%%%%%%%%%%%%%%%%%%%%%%%%%
\par
The up and down direction of the vibration can be obtained with (\ref{eqD2}).
%%%%%%%%%%%%%%%%%%%%%%%%%%%%%%%%%%%%%%%%%%%%%%%%%%%%%%%%%%%%%%%%%%
\begin{equation}
    \label{eqD2}
    Direction=\left\{ \begin{matrix}
	Up:&		\varDelta \cos A\,\,\leqslant 0 \&\& \varDelta \cos B\,\,\leqslant 0\\
	Down:&		\varDelta \cos A\,\,\geqslant 0 \&\& \varDelta \cos B\,\,\geqslant 0\\
\end{matrix} \right..
\end{equation}
%%%%%%%%%%%%%%%%%%%%%%%%%%%%%%%%%%%%%%%%%%%%%%%%%%%%%%%%%%%%%%%%%%
\section{Volume-based measurement}
Obviously, utilizing volume information enables the self-vibration measurement of a mmRadar swaying in spatial region. Following the same line of thought, at least three reference points are needed in space to form a simple spatial geometrical shape—a tetrahedron—with the radar. In three-dimensional space, the radar's swaying direction relative to the initial point can roughly be categorized into $2^3$ types. By leveraging the volume differences of the tetrahedron at different times, equations regarding the radar's self-vibration amplitude $r_k$ can be constructed. This facilitates the measurement of radar vibration amplitude and estimation of the quadrant-level vibration direction.
%%%%%%======================================%%%%%%%%%%%%%%%%
\section{Experiments}
\begin{figure*}[t]
\centering
%%%%%%%%%%%%%%%%%%%%%%%%%%
    \subfigure[\centering{X-swaying measurement results}]{\label{LR2d}\includegraphics[height=5.2cm]{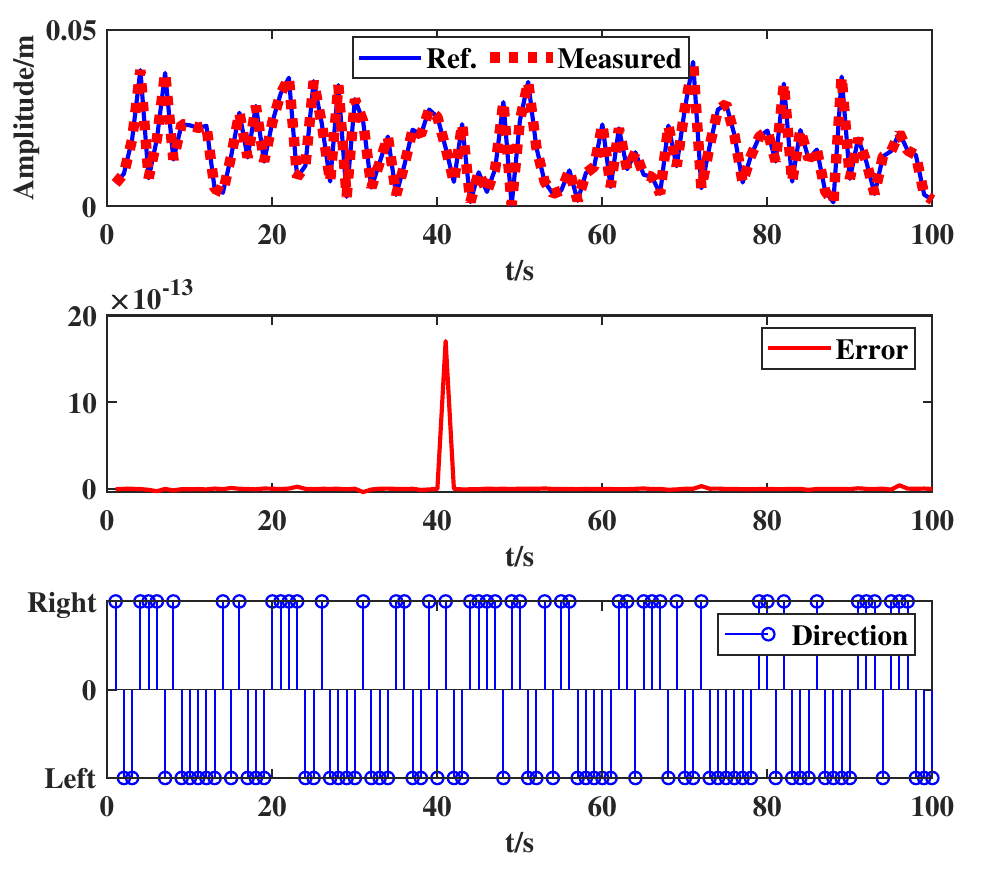}}
    \subfigure[\centering{Y-swaying measurement results}]{\label{UD2d}\includegraphics[height=5.2cm]{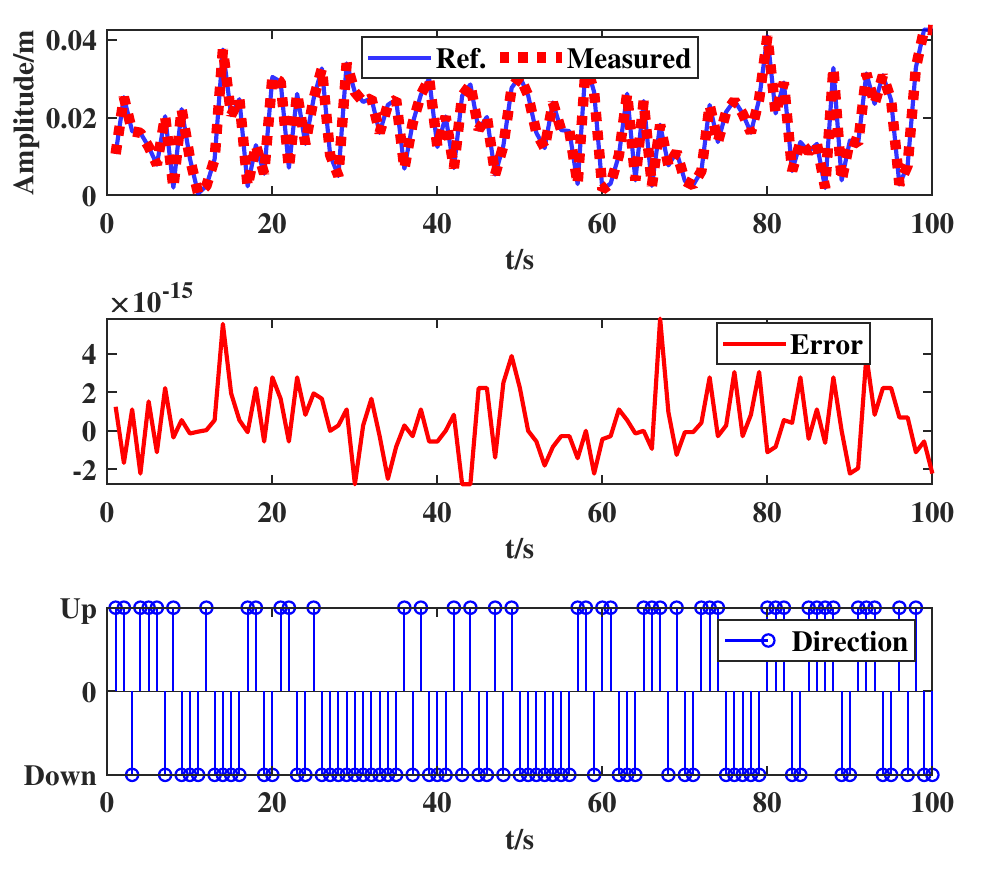}}
    \subfigure[\centering{XY-swaying measurement results}]{\label{LRUD2d}\includegraphics[height=5.2cm]{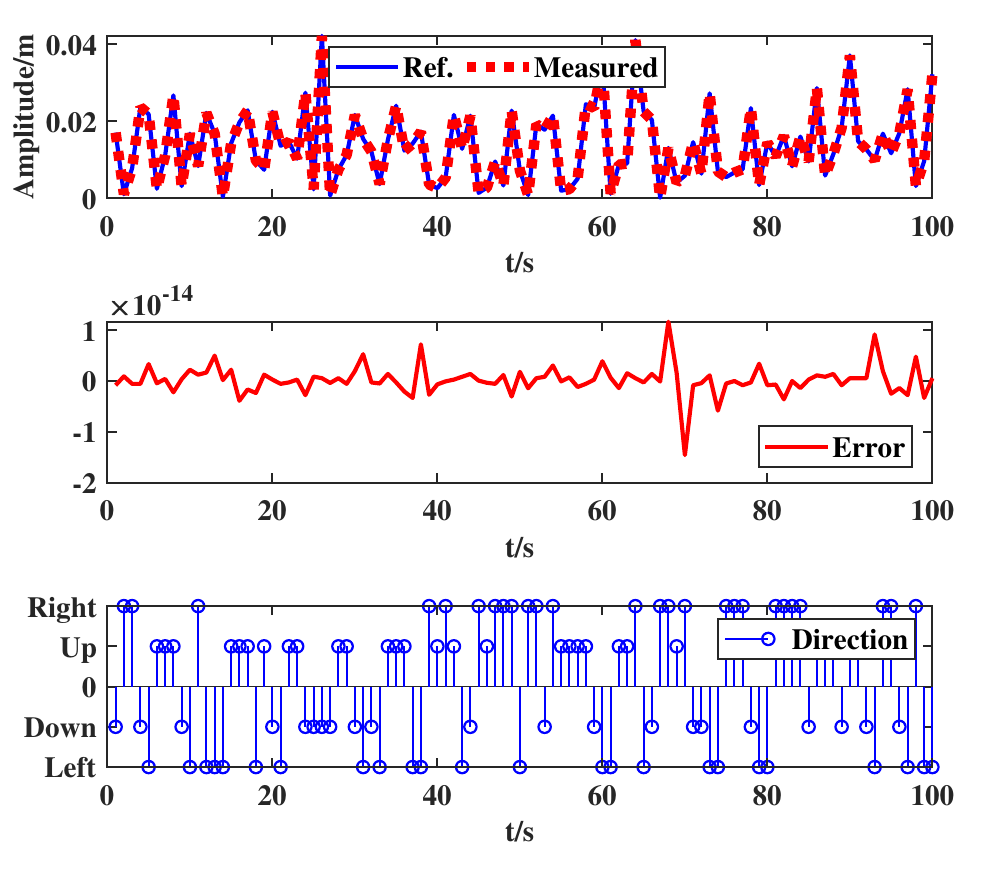}}\\
    \subfigure[\centering{X swaying of the radar}]{\label{LR3d}\includegraphics[height=6.0cm]{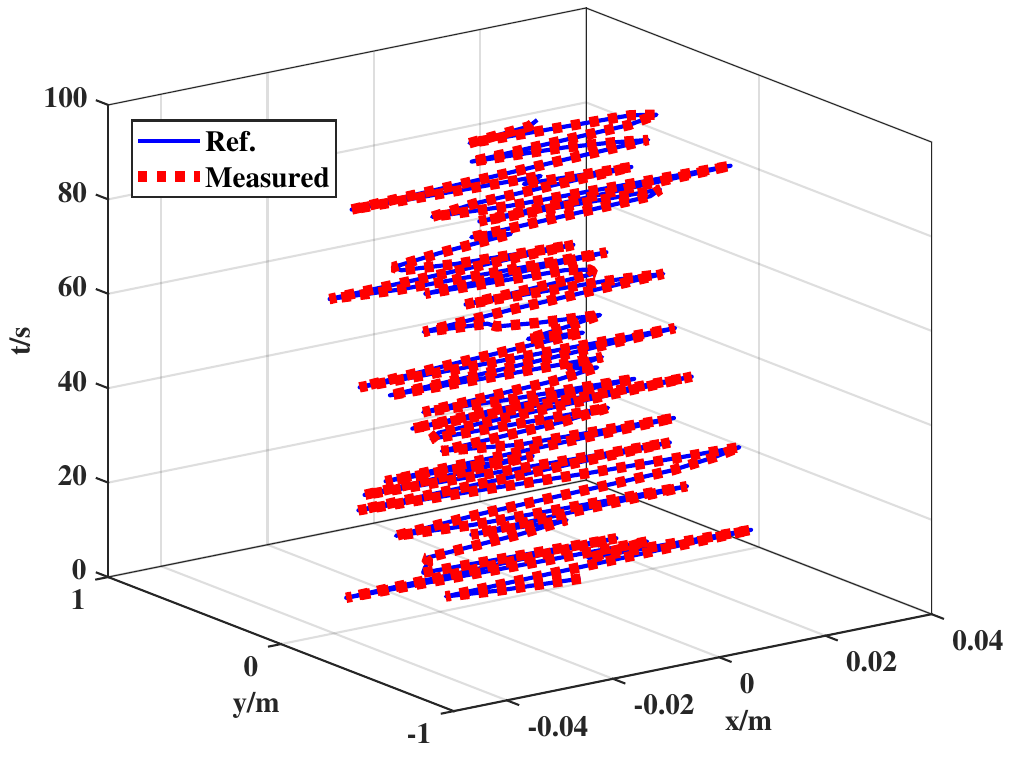}}
    \subfigure[\centering{Y swaying of the radar}]{\label{UD3d}\includegraphics[height=6.0cm]{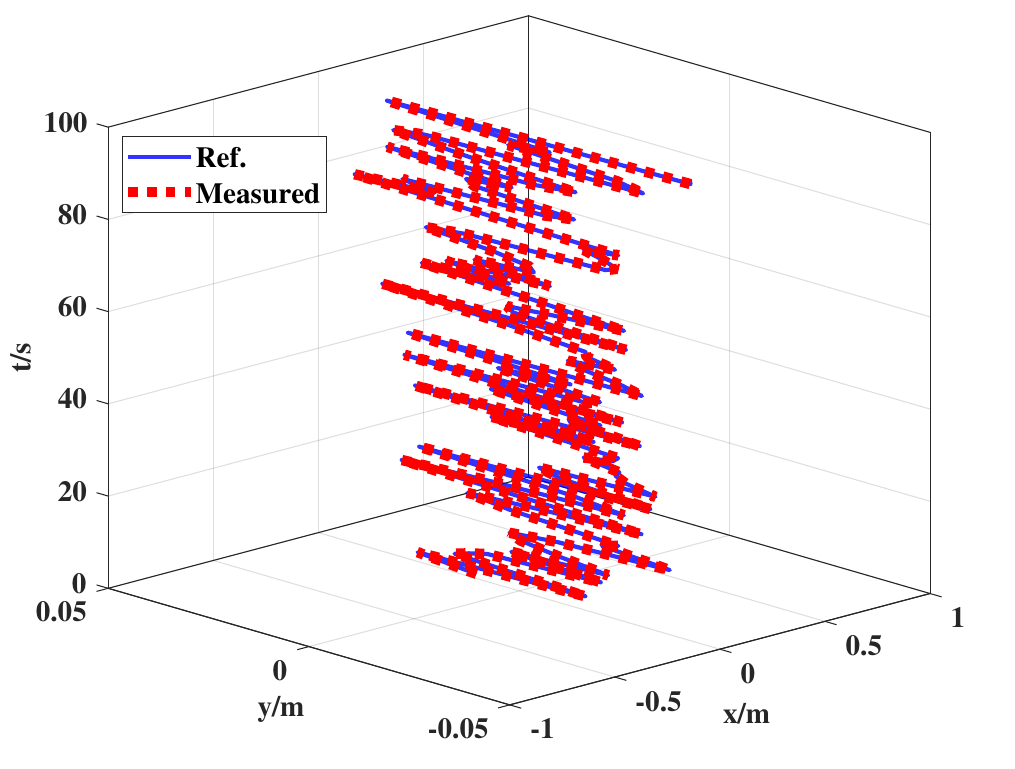}}\\
    \subfigure[\centering{XY swaying of the radar}]{\label{LRUD3d}\includegraphics[height=6.0cm]{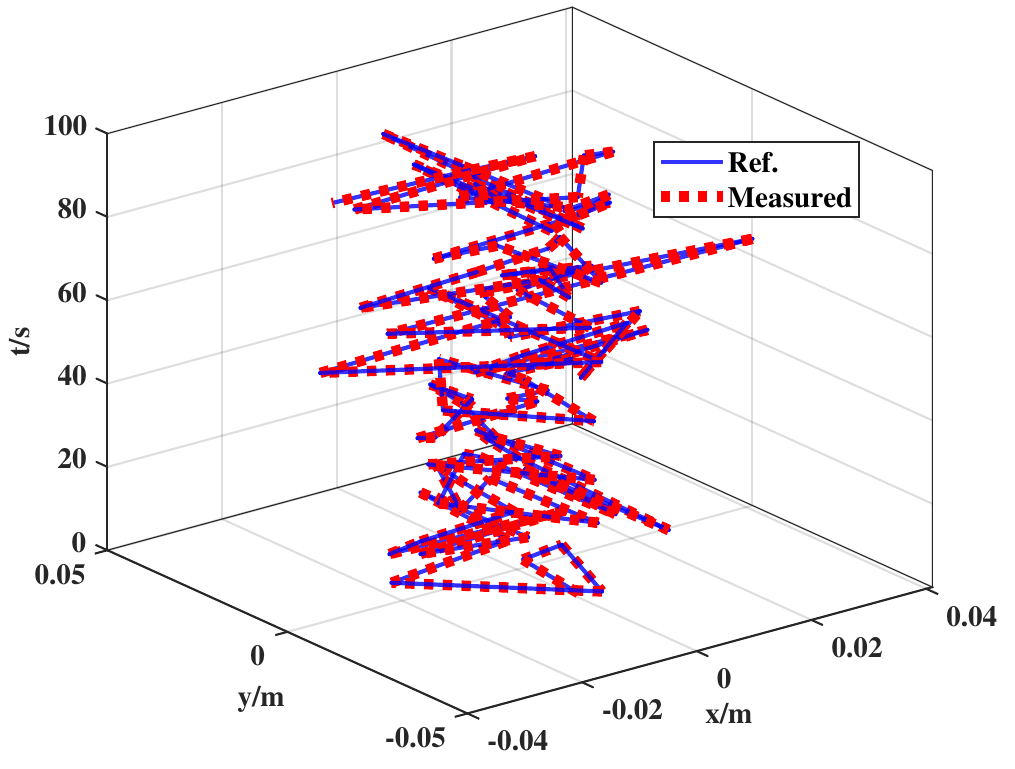}}
     \subfigure[\centering{Planar vibration of the XY swaying radar}]{\label{xy2d}\includegraphics[height=6.0cm]{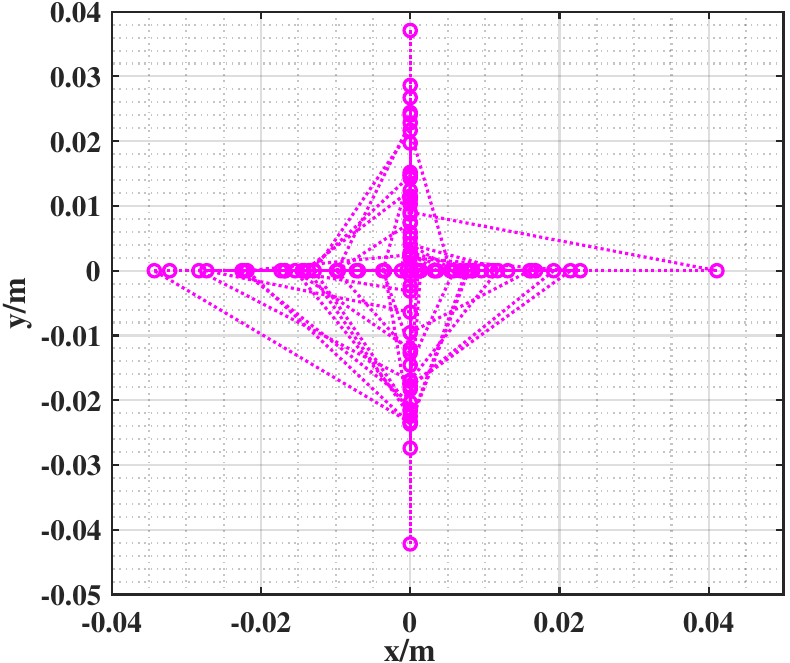}} 
%%%%%%%%%%%%%%%%%%%%%%%%%%
\caption{The measured results of planar vibrations of a swaying mmRadar.
\label{experiment}}
\end{figure*}

In this section, we show the results of using area information to measure the planar vibration of a swaying mmRadar, as shown in Fig. \ref{experiment}. From this, we can see that the radar can realize the measurement of its own vibration without additional sensors and only with the assistance of 2 references in the field of view.
%%%%%%======================================%%%%%%%%%%%%%%%%
%%%%%%======================================%%%%%%%%%%%%%%%%
%%%%%%%%%%%%%%%%%%%%%%%%%%%%%%%%%%%%%%%%%
\section{Conclusions}
This paper presents two new strategies for measuring the self-vibration of swaying radars using geometrical information. The methods are simple but effective, requiring only a single radar without additional sensors, thus avoiding complex synchronization operations and reducing the payload on mobile platforms. This study opens a new
window to measure radar self-vibration and may stimulate more ideas for airborne mmRadar-based wind detection, platform vibration fault signal analysis, and radar self-vibration cancellation algorithms.
\par
\textbf{Limitation and Future Work.}
The methodology proposed in this paper relies on the selection of reference objects in the radar operating environment. Due to the limited ranging, angular measurement, and spatial resolution capabilities of radar systems, the precision of reference object observations is constrained, thereby affecting the measurement accuracy of radar self-vibration. This limitation underscores the rationale for recommending the application of our methods to millimeter-wave radar. In the future, we would focus on integrating high-resolution algorithms to broaden the adaptability of the proposed methods.
%%%%%%%%%%%%%%%%%%%%%%%%%%%%%%%%%%%%%%%%%%%%%%%%%%%%%%%%%%
%%%%%%%%%%%%%%%%%%%%%%%%%%%%%%%%%%%%%%%%%
  %%%%%%%%%%%%%%%%%%%%%%%%%%%%%%%%%%
% Generated by IEEEtran.bst, version: 1.12 (2007/01/11)

  %%%%%%%%%%%%%%%%%%%%%%%%%%%%%%%%%%
  %%%%%%%%%%%%%%%%%%%%%%%%%%%%%%%%%%%%%%%%%%%%%%%%%%%%%%%%%%%%
\end{document}